# RIGID ROTATION AND THE KERR METRIC

Gerald E. Marsh

Argonne National Laboratory (Ret)
5433 East View Park
Chicago, IL 60615

E-mail: gemarsh@uchicago.edu

**Abstract.** The Einstein field equations have no known and acceptable interior solution that can be matched to an exterior Kerr field. In particular, there are no interior solutions that could represent objects like the Earth or other rigidly rotating astronomical bodies. It is shown here that there exist closed surfaces upon which the frame-dragging angular velocity and the red-shift factor for the Kerr metric are constant. These surfaces could serve as a boundary between rigidly rotating sources for the Kerr metric and the Kerr external field.

PACS: 04.20.Jb, 04.70.Bw

Key Words: Rigid Rotation; Kerr Metric

**Introduction.**

The Kerr solution to the Einstein field equations is generally thought to be the only possible stationary, axially symmetric and asymptotically flat solution that could represent the gravitational field outside an uncharged rotating body. It is characterized by two parameters, the angular momentum per unit mass *a*, and the mass *m*, and is often discussed in Boyer-Lindquist coordinates,[1,2] which will also be used here

The uniqueness of the Kerr solution does not mean that it plays the role of a Birkhoff theorem for rotating massive objects. What is the case is that the space-time geometry outside a rotating mass asymptotically approaches that of the Kerr solution. The reason for this is that the multipole moments of the Kerr solution are closely related while those of real mass distributions may in principle be independently specified. Because higher multipole fields fall off rapidly with distance from the source, the gravitational field of a rotating object will asymptotically approach that of the Kerr solution. There is then an apparent contradiction between the uniqueness theorems for the Kerr solution and the near external field of real rotating masses.

An approximation to the gravitational potential due to the multipoles of the Kerr solution[3] is given by $(-1)^{n+1} m \, (a^{2n}/r^{2n+1}) \, P_{2n}(\cos\theta)$, where *r* is the radial coordinate in Cartesian space. A real astronomical body undergoing gravitational collapse would have to selectively radiate away some of its multipole moments so as to satisfy this relation if the Kerr solution were to represent the end state of its exterior gravitational field. As put by Thorne[4] many years ago, "Because of this relationship between multipole moments and angular momentum, the Kerr solution cannot represent correctly the external field of *any* realistic stars (except for a <<set of measure zero>>)."

The other problem with the Kerr solution is that is has no known acceptable interior solution. That is, one that is non-singular and able to be matched to the exterior solution on the boundary; i.e., the metric tensor $g_{ij}$ and its first order partial derivatives should be continuous across the boundary[5,6] or, in the 3+1 formulation, $g_{ij}$ and the extrinsic curvature $K_{ij}$ must be continuous.[7]

**Interior Solutions.**

Because of the polar (or zenith) angle dependent frame-dragging effect inherent in the Kerr solution (also known as the Lense-Thirring effect), one generally considers some form of rotating fluid for the interior solution so as to be able to satisfy the boundary conditions, often on an oblate spheroidal coordinate surface corresponding to $r$ = constant in Boyer-Lindquist coordinates. Since the paper by Hernandez[3], a large literature on perfect fluid interior solutions for the Kerr metric has appeared. Krasinski[8] has given a careful review of the various approaches to this problem.

**Surfaces of constant red-shift factor and frame-dragging velocity**

Thorne's comment, quoted above, about the multipole moments of the Kerr solution does not rule out the existence of all interior solutions but only relegates the class of such solutions to "a set of measure zero" in the context of gravitational collapse. The introduction of surfaces of constant red-shift and frame-dragging velocity would simplify the problem of matching the Kerr exterior field to rotating solid bodies. The possibility of doing this was foreshadowed by Thorne.

In discussing the Kerr solution with regard to rotating objects, one generally considers only cases where $m > a$. But it should be noted, at least in passing, that most common rotating objects, like the Earth or a rotating 33 rpm record,[9] have parameters where $a \gg m$. For the vacuum Kerr solution this means the singularity is not hidden behind a horizon, but this would not be a problem for real rotating objects since the Kerr exterior solution would apply only outside the boundary of the interior solution for the object, and the interior solution would not be acceptable if it were singular.

The bounding surface of a rigidly rotating solid body has no differential rotation associated with it. If the Kerr field is to be matched to such a surface it must satisfy a number of conditions:

I. The exterior Kerr solution must have a closed surface such that on it the frame-dragging angular velocity is constant; i.e., independent of the Boyer-Lindquist $\theta$ coordinate.

II. That surface, if it exists, must also have a constant angular velocity as measured relative to a reference frame at infinity.

III. The red-shift factor, defined below, must also be a constant on the surface for photons emitted with zero angular momentum relative to the rotation axis.[4]

It will now be shown that it is possible to satisfy these three conditions, and that such surfaces do exist and can be found analytically. Several examples of such surfaces will be given. The parameters *a* and *m* will be used in each of these examples to find these surfaces and define the metric *outside* the surfaces. There is no intent or attempt made to actually match these surfaces to the boundaries of the examples.

The red-shift factor is defined as

$$\left(u^t\right)^{-2} = g_{tt} + 2g_{t\phi}\Omega + g_{\phi\phi}\Omega^2,$$

(1)

where $u^t$ is the time component of the 4-velocity and $\Omega = \dfrac{u^\phi}{u^t} = \dfrac{d\phi}{dt}$ is a the angular velocity as measured relative to a reference frame at infinity.

For a rigidly rotating body, the surface has no differential rotation so that $\Omega$ must be a constant. To match the Kerr exterior field, we also need to have $\omega$ constant so that there is no differential frame dragging at the boundary of the body. If we set these constants equal, we have $K = \omega = \Omega$. The red-shift factor then becomes

$$\left(u^t\right)^{-2} = g_{tt} - g_{t\phi}K.$$

(2)

For the Kerr solution one has

$$g_{tt} = \frac{r^2 - 2mr + a^2\cos^2\theta}{r^2 + a^2\cos^2\theta}, \quad g_{t\phi} = 2\frac{mra\sin^2\theta}{r^2 + a^2\cos^2\theta},$$

(3)

so that the red-shift factor becomes

$$(u^t)^{-2} = \frac{r^2 - 2mr + a^2\cos^2\theta - 2Kmra\sin^2\theta}{r^2 + a^2\cos^2\theta}.$$

(4)

Since this must be constant on the surface of the rigidly rotating body, we look for solutions to the quadratic equation

$$\frac{r^2 - 2mr + a^2\cos^2\theta - 2Kmra\sin^2\theta}{r^2 + a^2\cos^2\theta} - C = 0,$$

(5)

where $C$ is another constant not equal to $K$. This equation is soluble and yields the solutions

$$r_1 = -\frac{m + Kma\sin^2\theta + \left(\left(m + Kma\sin^2\theta\right)^2 - a^2(C-1)^2\cos^2\theta\right)^{1/2}}{C - 1},$$

$$r_2 = -\frac{m + Kma\sin^2\theta - \left(\left(m + Kma\sin^2\theta\right)^2 - a^2(C-1)^2\cos^2\theta\right)^{1/2}}{C - 1}.$$

(6)

The parameters and the range of $\theta$ must be such that

$$\left(\left(m + Kma\sin^2\theta\right)^2 - a^2(C-1)^2\cos^2\theta\right)^{1/2}$$

(7)

is real. Nonetheless, the values for which this expression is real will yield a complete surface. Note that $r_1$ and $r_2$ are invariant under the combination of $a \to -a$ and $K \to -K$. Such a transformation corresponds to a reversal of the direction of rotation.

The meaning of the constant $C$, which sets the value of constant red shift factor, can be understood by a comparison with the Newtonian potential, where $g_{tt} = 1-2U$, $g_{t\phi} = 0$, and $g_{\phi\phi} = -r^2\sin^2\theta$. Thus,

$$\left(u^t\right)^{-2} = 1 - 2U - r^2\Omega^2\sin^2\theta.$$

(8)

Here $U$ is the total Newtonian surface potential—with the positive sign convention, where $U > 0$. $C$ effectively controls the "radius" of the constant red-shift and frame-dragging surface. Radius is put in quotes since because the surface is not in general spherical.

Since $\left(u^t\right)^{-2}$ is a constant, rearranging the terms in Eq. (8) gives $U + \frac{1}{2}r^2\Omega^2\sin^2\theta = Constant$. This says that the total potential at the surface of a non-deformable rotating spherical body is constant. The second term on the left might be called the "centrifugal potential". In general, for a deformable or fluid body, the total potential $U$ includes the gravitational potential resulting from the change in radius at the surface due to the deformation as well as the potential due to the change in mass distribution caused by the deformation—sometimes called the "self-potential", both of which depend on $\theta$.

The red-shift factor $\left(u^t\right)^{-2}$ should not be confused with the actual red shift from the surface of the body, which is given by $1/u^t$. The distinction will be important in what follows.

In order to plot the solutions to Eqs. (6) in Cartesian coordinates one must convert the expressions given above in Boyer-Lindquist coordinates to Cartesian coordinates. The transformation from Boyer-Lindquist to Cartesian coordinates is

$$x = \left(r^2 + a^2\right)^{1/2}\sin\theta_{B-L}\cos\phi$$
$$y = \left(r^2 + a^2\right)^{1/2}\sin\theta_{B-L}\sin\phi$$
$$z = r\cos\theta_{B-L}.$$

(9)

For clarity, in the above equations and in what follows below, the subscripts *B-L* and *C* have been added where needed to distinguish between Boyer-Lindquist and Cartesian variables. From Eq. (9), one readily shows that

$$\frac{x^2 + y^2}{r^2 + a^2} + \frac{z^2}{r^2} = 1.$$

(10)

The relationship between the coordinates is shown in Figure 1, drawn for $\phi = 0$.

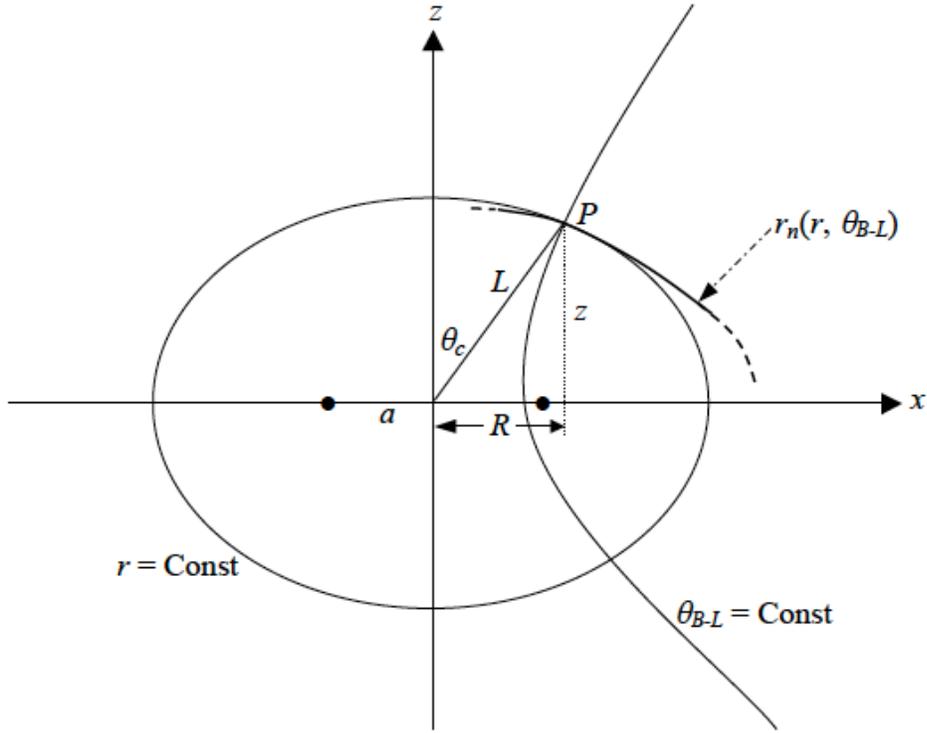

Figure 1. The figure shows a portion of the constant frame-dragging and red-shift surface given by $r_n(r,\theta_{B\text{-}L})$, where $n$ corresponds to one of the two roots $r_1$ or $r_2$ of Eq.(11). The constant Boyer-Lindquist coordinate surfaces that intersect $r_n(r,\theta_{B\text{-}L})$ at the point $P \in r_n(r,\theta_{B\text{-}L})$ are designated by $r$ = Const and $\theta_{B\text{-}L}$ = Const. $\theta_C$ is the Cartesian polar angle corresponding to the point $P$, which also has Cartesian coordinates $z$ and $x$. The distance $R = (x^2 + y^2)^{1/2}$, and the figure is drawn for $y = 0$ corresponding to $\phi = 0$. The two dots at $x = \pm a$ correspond to the ring singularity at $r = 0$, $\theta_{B\text{-}L} = \pi/2$.

From the figure, we have

$$z = L \cos \theta_C$$
$$R = L \sin \theta_C.$$

(11)

so that

$$\frac{L^2 \sin^2 \theta_C}{r^2 + a^2} + \frac{L^2 \cos^2 \theta_C}{r^2} = 1.$$

(12)

$L$ is then given by

$$L = \left( \frac{r^2(r^2 + a^2)}{r^2 + a^2 \cos^2 \theta_C} \right)^{1/2}.$$

(13)

With $\phi = 0$, Eqs. (11) and (13) allow a cross section of the constant frame-dragging and red-shift surfaces $r_n(r, \theta_{B-L})$ to be plotted in Cartesian coordinates. In doing so, however, the Boyer-Lindquist $\theta$ in $r_n$ will be interpreted by the plotting program of Mathematica®10 as $\theta_C$. It will be seen, however, that for the examples given below, the error is very small.

In order to plot the examples that follow, it is necessary to determine the value of the constant $C$. The red shift from a body of mass $m$ and radius $r$, as measured far from the body, is given in *mks* units by $1/u^t = \sqrt{1 - \frac{2Gm}{c^2 r}}$. $C$ is given by the square of this quantity. Three examples will be given, that of the Sun, the canonical neutron star (defined as having a radius of 10*km*, a mass of 1.4 solar masses, and a period of 1.5*ms*), and the Earth. In each of these examples, the "radius" of the constant red-shift and frame-dragging surface is greater than the positive horizon and ergosphere of the vacuum solution used to set the exterior field by a very comfortable margin, even in the case of the canonical neutron star.

The following table gives the value of $1/u^t$ for each of the three examples.

|  | $1/u^t = \sqrt{1 - \frac{2Gm}{c^2 r}}$ |
|---|---|
| SUN | 0.999997878 |
| NEUTRON STAR | 0.765871 |
| EARTH | 0.99999999932 |

Table 1. The red shift $1/u^t$ for the Sun, the canonical neutron star, and the Earth. Note that $C = (1/u^t)^2$.

The plot of the constant red-shift and frame-dragging surface $r_1$, for the Sun, is shown in Fig. 2. The oblateness is greatly exaggerated in the figure by the choice of aspect ratio; the actual eccentricity, given by

$$e = \sqrt{\frac{x^2 - z^2}{x^2}},$$

(14)

is essentially zero.

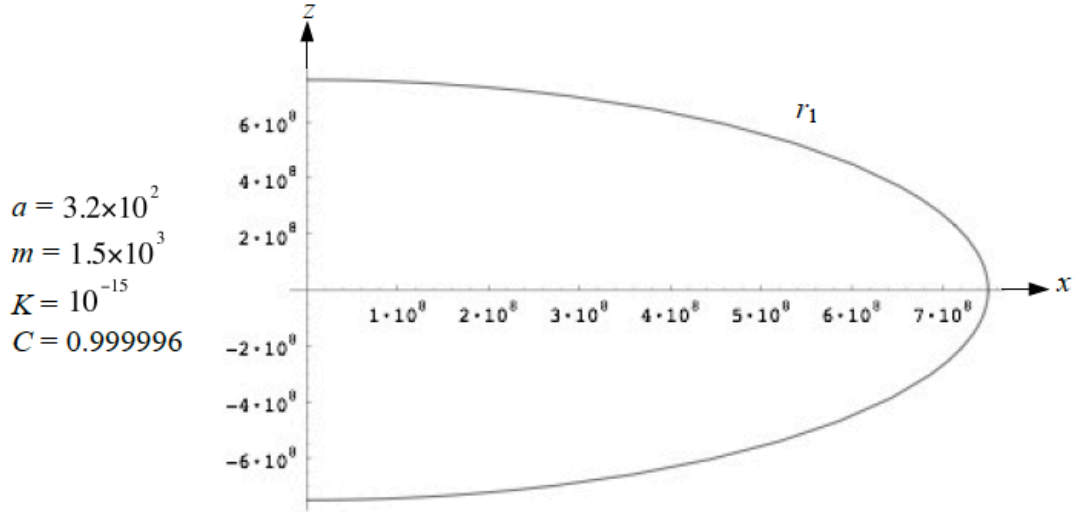

Figure 2. The constant red-shift and frame-dragging surface given by the first solution, $r_1$, of Eq.(11) for parameters corresponding to the Sun. $a$, $m$ and $K$ are in geometrized units, while $C$ is dimensionless. The oblateness of the surface is greatly exaggerated by the choice of aspect ratio. Because of the cylindrical symmetry, the full surface is obtained by rotating the figure around the $z$-axis.

For comparison, the radius of the Sun is $7 \times 10^8 m$, which is just slightly less than the radius of the constant red-shift and frame-dragging surface at $\theta = \pi/2$. The surface given by the second solution of Eq. (11), $r_2$, is not physically acceptable. The same is the case for $r_2$ of the other examples given below. The plotting method used for Fig. 2 also has large errors when applied to $r_2$. No further consideration will be given to this surface.

For the parameters corresponding to the canonical neutron star, one obtains the plot shown in Fig. 3. The value of $a$ is calculated from the neutron star angular momentum given by Dessart, et al.[11] and $K$ from the period.

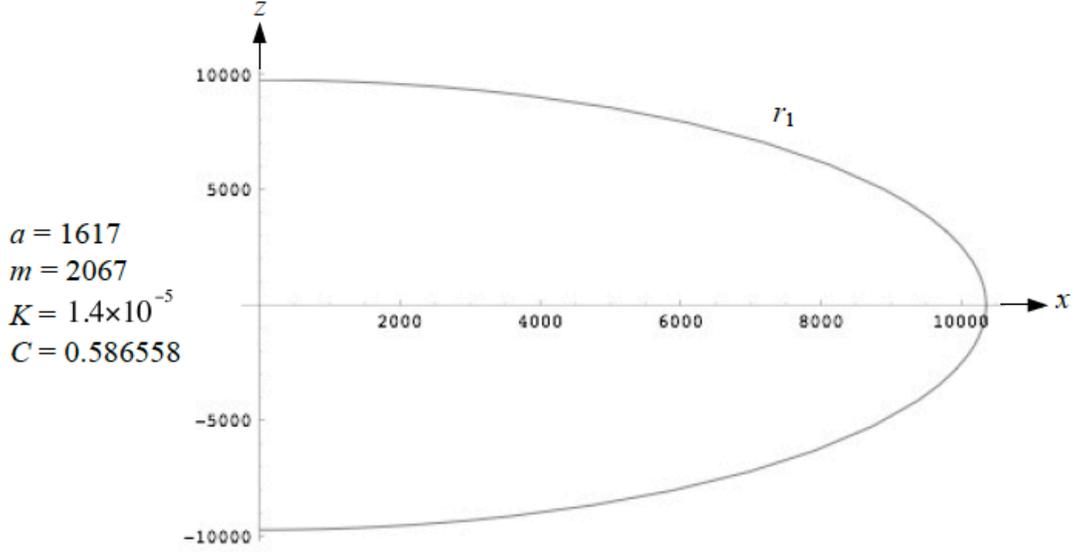

$a = 1617$
$m = 2067$
$K = 1.4 \times 10^{-5}$
$C = 0.586558$

Figure 3. Canonical neutron star constant red-shift and frame-dragging surface given by the first solution, $r_1$, of Eq.(11). $a$ is calculated from the neutron star angular momentum given by Dessart, et al. and $K$ from the period.

The eccentricity of the surface shown in Fig. 3 is $e = 0.34$. For this surface to be outside the neutron star, the eccentricity of the latter would have to be greater than this value. While it is close, this is likely to be the case.[12,13] The oblateness $\varepsilon$, determined from $\varepsilon + 1 = (1-e^2)^{-1/3}$, is 0.04, compared to the value of the Crab pulsar where $\varepsilon \sim 10^{-3}$, but the period of the Crab pulsar is 33ms compared to the 1.5ms of for this example. As can be seen, at $\theta = \pi/2$ the radius of this surface is 1.04 times greater than the canonical neutron star radius of 10 km.

The final example is that of the Earth, for which there is also data from the Gravity Probe B experiment. The frame-dragging measurement gave a magnitude of $37.2 \pm 7.2$ milliarc sec/yr or $\sim 6.4 \times 10^{-15}$ rad/sec. Converted to geometrical units, this is $\sim 2.1 \times 10^{-23}$ $m^{-1}$. The surface is shown in Fig. 4.

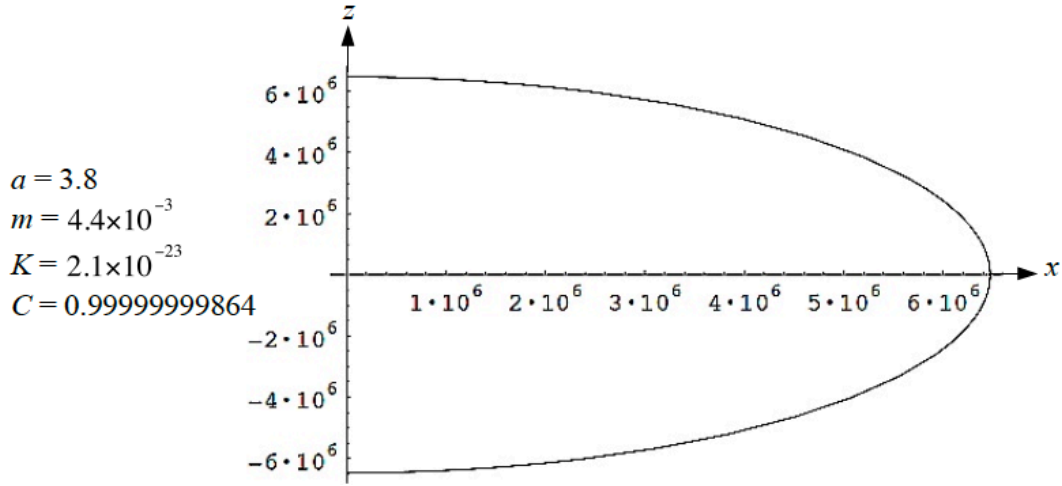

Figure 4. Constant red-shift and frame-dragging surface given by the first solution, $r_1$, of Eq.(11) for the Earth. The constant $K$ is set by the frame–dragging measurements of the Gravity probe B experiment. Note that for the Earth $a \gg m$.

Here the eccentricity is, as was the case for the Sun, essentially zero. Since the radius of the Earth is $6.4 \times 10^6$ m, at $\theta = \pi/2$ this surface, with radius $\sim 6.48 \times 10^6$, is only slightly larger than the radius of the Earth., and just a bit smaller than the Gravity Probe B orbital radius of $\sim 7 \times 10^6$.

**The multipole issue**

The magnitude of the multipole contribution to the potential at the location of the surfaces of constant red shift and frame dragging will now be shown to be very small compared to the Newtonian potential.

The approximate potential for the multipoles associated with the Kerr metric that was given by Hernandez, Jr. and discussed above can be written as

$$V = -\sum_{n=0}^{\infty} \frac{(-1)^n m\, a^{2n}}{r^{2n+1}} P_{2n}(\cos\theta) = -\frac{m}{r} - \sum_{n=1}^{\infty} \frac{(-1)^n m\, a^{2n}}{r^{2n+1}} P_{2n}(\cos\theta).$$

(15)

The first term on the right hand side of this equation is, of course, the Newtonian potential, while the second is the contribution of the multipoles.

The structure of this potential is not unique to the Kerr metric. It also appears when computing the Newtonian potential for an oblate spheroid[14], whose ellipticity is not too great, at points exterior to the spheroid. The above examples meet this criterion, and the potential can be shown to be

$$V_{Newtonian} = -\frac{m}{r} - \sum_{n=1}^{\infty} \frac{(-1)^n m\, a^{2n}}{(2n+1)(2n+3)\, r^{2n+1}} P_{2n}(\cos\theta).$$

(16)

In the Newtonian case, the multipole fields fall off faster because of the numerical factor $(2n+1)(2n+3)$ in the denominator.

For the examples above, Table 2 shows the magnitude of contribution to the potential of the first 10 terms of the multipole expansion of Eq. (15) compared to that of the Newtonian potential. The nominal radius and eccentricity are also given for comparison purposes.

|  | Nominal Radius (m) | Eccentricity | $m/z$ ($\theta = 0$) | $m/x$ ($\theta = \pi/2$) | Multipole |
|---|---|---|---|---|---|
| SUN | $7\times10^8$ | ~0 | $2\times10^{-6}$ | $2\times10^{-6}$ | $2\times10^{-19}$ |
| NEUTRON STAR | $10^4$ | 0.34 | 0.212 | 0.199 | $5.71\times10^{-3}$ ($\theta=0$) <br> $2.5\times10^{-3}$ ($\theta=\pi/2$) |
| EARTH | $6\times10^6$ | ~0 | $6.8\times10^{-10}$ | $6.8\times10^{-10}$ | $10^{-22}$ |

Table 1. The relative contributions to the overall potential of the first ten terms of the multipole expansion compared to that of the Newtonian potential at the position of the constant red-shift and frame-dragging surface. Note that the $r$ in Eq. (15) is replaced with the $L$ of Eq. (13) for calculating the entries in the last column of the table.

As can be seen, even in the case of the neutron star, the multipole contribution to the potential at the position of the constant red-shift and frame-dragging surface is very small compared to the Newtonian potential. For the neutron star, the only case for which there is a significant difference, the potential due to the multipoles is given for the $z$-axis at $\theta = 0$ and the $x$-axis at $\theta = \pi/2$.

**Plotting errors**

The plotting error for Figs. 2, 3, and 4 can be estimated from

$$\cos\theta_C - \cos\theta_{B-L} = \frac{z}{L} - \frac{z}{r} = \left(\frac{1}{L} - \frac{1}{r}\right)z = \left(\frac{1}{L} - \frac{1}{r}\right)L\cos\theta_C.$$

(17)

Plotting this expression for the neutron star, which has the largest error—the other examples having an error on the order of a few times $10^{-14}$, gives the result shown in Fig. 5.

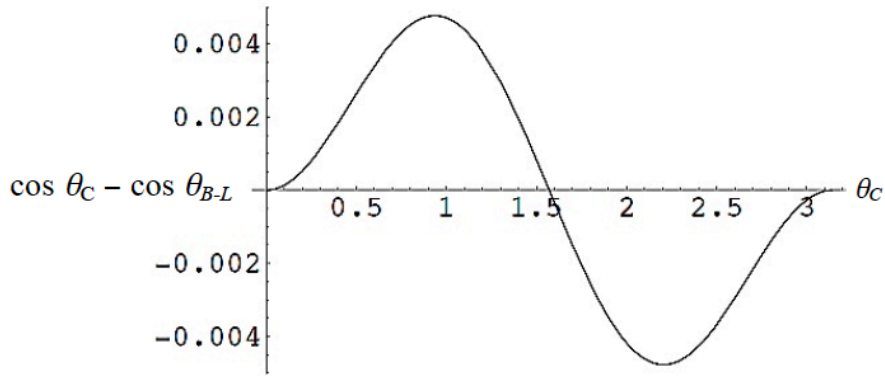

Figure 5. The plotting error introduced by the method of plotting used in the figures above is greatest for the case of the neutron star and is shown in this figure.

Note the plotting error is very small and vanishes for $\theta = 0$, $\pi/2$, and $\pi$.

**Summary**

The differential frame-dragging effect inherent in the Kerr metric generally restricts consideration to some form of rotating fluid for the interior solution so as to be able to satisfy the boundary conditions. However, it has been shown here that there exist surfaces of constant red-shift and frame-dragging angular velocity that could serve as the boundary between the exterior Kerr field and an interior solution for a rigidly rotating solid body. Examples of such surfaces were found for parameters corresponding to the Sun, the canonical neutron star, and the Earth. The results are at least consistent with actual data from neutron star modeling and the Gravity Probe B experiment.

**REFERENCES**

[1] R. H. Boyer and R. W. Lindquist, "Maximal Analytic Extension of the Kerr Metric", *J. Math. Phys.* **8**, 265-281 (1967).

[2] Geometrical units are used; see, for example: R.M. Wald, *General Relativity*, Appendix F (University of Chicago Press, Chicago 1984). For convenience, the general connection between conventional and geometrical coordinates is as follows: If a quantity in *mks* units has the dimension $L^n T^m M^p$, where *L*, *T*, *M*, correspond to length, time, and mass respectively, then in geometrical units the dimension would be given by $L^{n+m+p}$. The conversion factor from *mks* to geometric units is . $c^m (c^2/G)^p$, where *c* is the velocity of light and G is the gravitational constant.

[3] W.C. Hernandez, Jr., "Material Sources for the Kerr Metric", *Phys. Rev.* **159**, 1070-1072 (1967).

[4] K.S. Thorne, "Relativistic Stars, Black Holes and Gravitational Waves (Including an in-Depth Review of the Theory of Rotating, Relativistic Stars), contained in B.K. Sachs, ed., *General Relativity and Cosmology*, Proceedings of the International School of Physics <<Enrico Fermi>> Course XLVII, pp. 237-283, Academic Press (1971).

[5] E.H. Robson, "Junction conditions in general relativity theory", *Annales de l'I.H.P.*, Section A, tome 16, n°1 (1972), pp. 41-5.

[6] J.L. Synge, *Relativity-The General Theory* (North-Holland Publishing Co., Amsterdam 1966), §9.

[7] C.W. Misner, K.S. Thorne, and J.A. Wheeler, *Gravitation* (W.H. Freeman and Co., San Francisco 1973), §21.13.

[8] A. Krasinski, "Ellipsoidal Space-Times, Sources for the Kerr Metric", *Annals of Physics* **112**, 22-40 (1978).

[9] C. Hoenselaers, "The Double Kerr Solution: A Survey", *Proceedings of the Fourth Marcel Grossmann Meeting on General Relativity*, R. Ruffini (ed), pp. 967-975 (1986).

[10] In Mathematica® one must also take into account that $\theta = 0$ corresponds to the *x*-axis while in B-L coordinates it is the *z*-axis.

[11] L. Dessart, et al., "Multi-Dimensional Simulations of the Accretion-Induced Collapse of White Dwarfs to Neutron Stars", *ApJ* **644**, 1063 (2006); (ArXiv:astro-phy0601603). (2006).

[12] S.L. Shapiro and S.A. Teukolsky, *Black Holes, White Dwarfs and Neutron Stars: The Physics of Compact Objects* (Wiley-VCH Verlag GmbH & Co. KGaA, Weinheim 2004).

[13] L.S. Finn and S.L. Shapiro, *ApJ* **459**, 444 (1990).

[14] A.G. Webster, *The dynamics of particles and of rigid, elastic, and fluid bodies* (B.G. Teubner, Leipzig 1904), Ch. VIII, Art.161, p. 424.